\documentclass[fleqn,usenatbib]{mnras}
\DeclareMathAlphabet{\mathcal}{OMS}{cmsy}{m}{n}
\DeclareMathAlphabet\mathbfcal{OMS}{cmsy}{b}{n}

\usepackage[T1]{fontenc}
\usepackage{graphicx}	% Including figure files
\usepackage{amsmath}	% Advanced maths commands
\usepackage{amssymb}	% Extra maths symbols
\usepackage{bm}
\usepackage{ulem}
\usepackage[dvipsnames]{xcolor}
\usepackage{tablefootnote}

\usepackage{chngpage}
\usepackage{multirow}
\usepackage{tabularx}
\usepackage{soul}

\usepackage[utf8]{inputenc}

\newcommand{\cosmosis}{\texttt{CosmoSIS}}
\newcommand{\class}{\texttt{CLASS}}
\newcommand{\emcee}{\texttt{emcee}}

\newcommand{\camb}{\texttt{CAMB}}
\newcommand{\lcdm}{$\Lambda$CDM}

\newcommand{\de}{\mathrm{d}}
\newcommand{\ho}{H_0}
\newcommand{\om}{\Omega_{\rm m}}
\newcommand{\ob}{\Omega_{\rm b}}
\newcommand{\ns}{n_{\rm s}}
\newcommand{\As}{A_{\rm s}}

\newcommand{\Q}{\mathcal Q}

\defcitealias{Tanidis:2019teo}{Paper I}
\defcitealias{Tanidis:2019mag}{Paper II}

\title[Unified LSS data analysis pipeline III: Multi-tracing]{Developing a unified pipeline for large-scale structure data analysis with angular power spectra -- III. Implementing the multi-tracer technique to constrain neutrino masses}
\author[K.\ Tanidis \& S.\ Camera]{Konstantinos Tanidis$^{1,2}$\thanks{tanidis@to.infn.it} and Stefano Camera$^{1,2,3,4}$\thanks{stefano.camera@unito.it}\\
$^{1}$Dipartimento di Fisica, Universit\`a degli Studi di Torino, via P.\ Giuria 1, 10125 Torino, Italy\\
$^{2}$INFN -- Istituto Nazionale di Fisica Nucleare, Sezione di Torino, via P.\ Giuria 1, 10125 Torino, Italy\\
$^{3}$INAF -- Istituto Nazionale di Astrofisica, Osservatorio Astrofisico di Torino, strada Osservatorio 20, 10025 Pino Torinese, Italy\\
$^{4}$Department of Physics \& Astronomy, University of the Western Cape, Cape Town 7535, South Africa
}
\date{Accepted XXX. Received YYY; in original form ZZZ}
\pubyear{2019}

\begin{document}
\label{firstpage}
\pagerange{\pageref{firstpage}--\pageref{lastpage}}
\maketitle

\begin{abstract}
In this paper, we apply the multi-tracer technique to harmonic-space (i.e.\ angular) power spectra with a likelihood-based approach. This goes beyond the usual Fisher matrix formalism hitherto implemented in forecasts with angular statistics, opening up a window for future developments and direct application to available data sets. We also release a fully-operational modified version of the publicly available code \cosmosis, where we consistently include all the add-ons presented in the previous papers of this series. The result is a modular cosmological parameter estimation suite for angular power spectra of galaxy number counts, allowing for single and multiple tracers, and including density fluctuations, redshift-space distortions, and weak lensing magnification. We demonstrate the improvement on parameter constraints enabled by the use of multiple tracers on a multi-tracing analysis of luminous red galaxies and emission line galaxies. We obtain an enhancement of $44\%$ on the $2\sigma$ upper bound on the sum of neutrino masses. Our code is publicly available at \url{https://github.com/ktanidis/Modified_CosmoSIS_for_galaxy_number_count_angular_power_spectra}.
\end{abstract}


%%%%%%%%%%%%%%%%%%%%%%%%%%%%%%%%%%%%%%%%%%%%%%%%%%

%%%%%%%%%%%%%%%%% BODY OF PAPER %%%%%%%%%%%%%%%%%%

\section{Introduction}
The present article is the third of a series of papers focussed on the set up of a pipeline to analyse consistently large-scale structure data sets through \textit{angular power spectra}---namely, two-point summary statistics of sky maps of a given observable, decomposed in harmonic space. This approach, which falls within the category of so-called forward modelling, is directly related to how observations are made and is inherently gauge-independent. Furthermore, it allows us to analyse local and non-local probes within the same framework, thus making profit of both auto- and cross-correlations, and avoiding the risk of underestimating their covariance \citep[see also][]{Krause:2016jvl,Nicola:2017ryw,Chisari_2019,Fang_2020,Joachimi:2020abi}.

The previous papers of our series are \citet[][hereafter `Paper I']{Tanidis:2019teo} and \citet[][`Paper II']{Tanidis:2019mag}, where we respectively studied the effect of redshift-space distortions (RSD) or magnification bias as corrections to the angular power spectrum of galaxy number counts. Specifically, we quantified the effect of neglecting such corrections on cosmological parameter estimation, and we showed how to include them in a fast, likelihood-based pipeline. We adopted the Limber approximation, focussed on the linear regime, and made use of for future large-scale structure surveys as benchmarks for our analyses.

In the present work, we capitalise on that and implement in our pipeline the `multi-tracer technique' \citep{Selijak2009,McDonald_2009}. Thanks to it, we can effectively measure the ratio of the biases of two or more tracers of the cosmic structure \citep[see e.g.][]{Witzemann:2018cdx}, net of the stochastic nature of the underlying density fluctuations. As a case study, we choose to investigate how constraints on the sum of the neutrino masses improve with a multi-tracing analysis. To do so, we generalise our method to allow for scale dependence in the growth factor and rate as well as in the bias, as expected in massive-neutrino cosmologies \citep{LESGOURGUES2006307}. At the same time, we make our fully-operational modified version of the \cosmosis\ code \citep{Zuntz2015} public.\footnote{\url{https://github.com/ktanidis/Modified_CosmoSIS_for_galaxy_number_count_angular_power_spectra}.} Our code accounts for RSD, the magnification bias effect, multi-tracing, and the aforementioned scale dependence for fundamental quantities like growth and bias, with the possibility for the user to switch on and off these effects at will.

The paper is outlined as follows. In \autoref{sec:formalism}, we extend the Limber approximated angular power spectra to the case of multi-tracing and define the RSD and magnification bias scale-dependent corrections. In \autoref{sec:neutrinos}, we review effect on the linear galaxy bias and growth of massive neutrinos, which we adopt as case study. In \autoref{sec:thesurvey}, we present the technical specifications of our benchmark survey, the Dark Energy Spectroscopic Instrument \citep[DESI][]{DESIcollab2016}. Then, in \autoref{sec:likelihood_scale_cuts} we show the likelihood used for the forecast as well as the multipole cuts applied in our analysis. Finally, results are presented in \autoref{sec:results} and concluding remarks are drawn in \autoref{sec:conclusions}. The code is described in appendix~\ref{appendix}.

%ArXiv:1911.11947

\section{Multi-Tracer Angular Power Spectra}
\label{sec:formalism}
The multi-tracer technique was first proposed in \citet{Selijak2009} as a means to overcome cosmic variance when studying the clustering of biased tracers of the large-scale structure \citep[see also][]{Abramo_2013,Abramo_2015}. The idea stems from the fact that quantities like bias and growth are, at least on large scales, deterministic and hence not affected by cosmic variance. The technique makes use of two or more different galaxy populations, which, by definition, are not independent since they all trace the same underlying matter field, but with different biases \citep{Guzzo_1997,Benoist_1996}. Thus, by multi-tracing, we effectively measure the ratio of the biases \citep[see e.g.][]{Witzemann:2018cdx}, two at a time, net of the stochastic nature of the power spectrum. Thanks to the sensitivity of multi-tracing to scale dependence, it was also suggested as a tool to measure the growth of structure from RSD \citep{McDonald_2009,viljoen2020constraining}.
 
As mentioned above, we focus on harmonic-space power spectra of galaxy number counts. Let us start with a set of:
\begin{itemize}
    \item $N_{\rm tr}$ tracers of the underlying large-scale cosmic structure (different galaxy populations, in the present case), labelled by upper-case Latin letters from the beginning of the alphabet, viz.\ $A,B\ldots$;
    \item $N_z^A$ redshift bins for each tracer, labelled by lower-case Latin letters from the middle of the alphabet, e.g.\ $i,j\ldots$
    % \item $N_\ell$ multipole bins;
\end{itemize}
Now, the harmonic-space power spectrum of the clustering between $A$-type galaxies in redshift bin $i$ and $B$-type galaxies in redshift bin $j$, $C^{\rm g}_\ell(z^A_i,z^B_j)$, is implicitly defined as \citep{Ferramacho:2014pua,Fonseca_2015}
\begin{equation}
    \left\langle g^{A,i}_{\ell m} g^{B,j\ast}_{\ell^\prime m'}\right\rangle=\delta^{\rm K}_{\ell\ell^\prime}\delta^{\rm K}_{m m'}C^{\rm g}_\ell(z^A_i,z^B_j),
\end{equation}
where e.g.\ $g^{A,i}_{\ell m}$ stands for the spherical-harmonic expansion coefficients of the sky map of the distribution of the $A$th tracer in the $i$th redshift bin, $\ast$ means complex conjugation, angle brackets denote ensemble average, an asterisk means complex-conjugation, and $\delta^{\rm K}$ is the Kronecker symbol. Clearly, the case $A=B$ reduces to the standard single-tracer analysis, whereas $i=j$ means restricting to auto-bin correlations.

If we focus on the main contributions to galaxy number density fluctuations---namely matter density perturbations, RSD, and magnification \citep[see e.g.][]{Yoo2010,ChallinorLewis2011,BonvinDurrer2011}---the theoretical expectation of the harmonic-space power spectrum in a given cosmology can be computed in the Limber approximation at $\ell\gg1$ via
\begin{equation}
    C^{\rm g}_{\ell\gg1}(z^A_i,z^B_j)=\int\frac{\de\chi}{\chi^2}\,W^{A,i}_{\rm g}(k_\ell,\chi)W^{B,j}_{\rm g}(k_\ell,\chi)P_{\rm lin}\left(k_\ell \right),
    \label{eq:Cl_TOT}
\end{equation}
with $k_\ell=(\ell+1/2)/\chi$ and $\chi$ the radial comoving distance to redshift $z$, whilst
\begin{equation}
    W^{A,i}_{\rm g}(k_\ell,\chi)=W^{A,i}_{\rm g,den}(k_\ell,\chi)+W^{A,i}_{\rm g,RSD}(k_\ell,\chi)+W^{A,i}_{\rm g,mag}(k_\ell,\chi)
    \label{eq:W_TOT}
\end{equation}
is the total weight function for each galaxy population and redshift-bin pair. The three terms of \autoref{eq:W_TOT} respectively read:
\begin{equation}
    W^{A,i}_{\rm g,den}(k_\ell,\chi)=n_i^A(\chi)b_i^A(k_\ell,\chi)D(k_\ell,\chi),
    \label{eq:W_den}
\end{equation}
for matter density fluctuations;
\begin{multline}
    W^{A,i}_{\rm g,RSD}(k_\ell,\chi)=\frac{2\ell^2+2\ell-1}{(2\ell-1)(2\ell+3)}n_i^A(\chi)\left[fD\right](k_\ell,\chi)\\ 
    -\frac{(\ell-1)\ell}{(2\ell-1)\sqrt{(2\ell-3)(2\ell+1)}}n_i^A \left(\frac{2\ell-3}{2\ell+1}\chi\right)\left[ fD\right]\left(k_\ell,\frac{2\ell-3}{2\ell+1}\chi\right)\\
    -\frac{(\ell+1)(\ell+2)}{(2\ell+3)\sqrt{(2\ell+1)(2\ell+5)}}n_i^A\left(\frac{2\ell+5}{2\ell+1}\chi\right) \left[fD\right]\left(k_\ell,\frac{2\ell+5}{2\ell+1}\chi\right),
    \label{eq:W_RSD}
\end{multline}
for RSD; and
\begin{multline}
    W^{A,i}_{\rm g,mag}(k_\ell,\chi)=\\
    3\om\ho^2\left[1+z(\chi)\right]\chi\tilde n_i^A(\chi)\left[\Q^A(\chi)-1\right]D(k_\ell,\chi),
    \label{eq:W_mag}
\end{multline}
for magnification.\footnote{In \citet[][]{Chisari_2019}, Limber-approximated angular power spectra are computed integrating over $k$ after the transformation $\chi_\ell=(\ell+1/2)/k$ has been performed, thus leading to different expressions for the weight functions.}

In \autoref{eq:Cl_TOT} to \autoref{eq:W_mag}, we have introduced the following quantities: the present-day linear matter power spectrum, $P_{\rm lin}(k)\equiv P_{\rm lin}(k,z=0)$, provided by \camb\ \citep{LCL2000}; the linear galaxy bias, $b(k,z)$; the growth factor, defined as $D(k,z)=\sqrt{P_{\rm lin}(k,z)/P_{\rm lin}(k)}$; the growth rate of matter perturbations, $f(k,z)=-(1+z) \de\ln{T(k,z)}/\de z$, with $T(k,z)$ the transfer function; and the source redshift distribution of the $A$th galaxy population in the $i$th redshift bin, $n_i^A(z)$.\footnote{Note that for any source distribution $n(z)\,\de z=n(\chi)\,\de\chi$ holds.} Moreover, in \autoref{eq:W_mag}, $\om$ and $H_0$ are respectively the total matter fraction in the Universe and the Hubble constant at present, we use units such that the speed of light $c=1$, whilst $\mathcal Q^A=5s^A/2$ is the magnification bias for tracer $A$, with $s^A$ the slope of the decadic logarithm of the \textit{comoving} galaxy number density as a function of observed magnitude, taken at the magnitude cut of the survey. Lastly, we have defined
\begin{equation}
\tilde n_i^A(\chi)=\int_\chi^\infty\de\chi'\,\frac{\chi'-\chi}{\chi'}n_i^A(\chi').
\end{equation}
This quantity is often called the lensing efficiency.

\section{Neutrinos free streaming length and scale dependent galaxy bias}
\label{sec:neutrinos}
Since the multi-tracer technique is best suited to detect the scale dependence of quantities such as the growth and the bias, we choose to study massive neutrinos, which are known to have such peculiarities. In this section, we shall review their basic characteristics relevant for our work.

Cosmic neutrinos are the most abundant particles in the Universe after photons, yet little is known about their mass and energy budget contribution today. Nonetheless, it has been detected with high significance that there are three neutrino species each having a radiation number density contribution $\approx 112/$cm$^3$ in the early Universe and a temperature around $\approx 1.6\times10^{-4}\,\mathrm{eV}$ \citep{Ade2015}. In addition to this, the neutrino oscillation measurements have shown that there are at least two massive out of the three neutrino mass eigenstates, contrary to the Standard Model of Particle Physics describing them as as fundamental but massless particles, and therefore suggesting that there is evidence for physics beyond the Standard Model.

Neutrino oscillation measurements provide us not with the individual neutrino masses, but rather with the two squared mass splittings. The relic neutrino density along with the oscillation data gives us a lower limit on the sum of the masses of the three neutrino species, namely $\sum m_\nu \ge 0.06\,\mathrm{eV}$ for what is known as \textit{normal} hierarchy, and $\ge 0.1\,\mathrm{eV}$ for \textit{inverted} hierarchy, whilst the upper limit provided by the latest cosmological data suggests $\sum m_\nu \le 0.15\,\mathrm{eV}$ at $95\%$ C.L.\ \citep{Ade2015,CAPOZZI2016218}.

Generally, the presence of massive neutrinos affects the total matter fraction in the Universe in the sense that the linear growth of the matter fluctuations is suppressed for scales $k<k_{\rm fs}$, with $k_{\rm fs}$ the neutrinos free streaming scale \citep{LESGOURGUES2006307}. This effect can be studied with lensing of the cosmic microwave background, the clustering of galaxies and other biased tracers of the large scale structure, and the weak lensing effect of cosmic shear, thus putting constraints on the sum of neutrino masses \citep{PhysRevLett.80.5255}. Another consequence of the presence of massive neutrinos is a scale dependence induced on the linear growth rate of cosmic structures and on the galaxy bias \citep{LoVerde2014,Font_Ribera_2014,upadhye2016}.

The comoving free-streaming scale is a redshift dependent quantity defined as
\begin{equation}
    k_{\rm fs}(z)=\sqrt{1.5} \frac{H(z)}{u_{\rm th} (1+z)},
    \label{eqn:free_stream_wave1}
\end{equation}
with the $H(z)$ the Hubble parameter and $u_{\rm th}$ the neutrino thermal velocity. When neutrinos are relativistic, their $k_{\rm fs}$ decreases as in \autoref{eqn:free_stream_wave1}. However, after neutrinos become non-relativistic, their thermal velocity starts to decay as
\begin{align}
    u_{\rm th} &\approx \frac{3 T_\nu}{m_\nu}\\
    &= 3 (4/11)^{1/3}\frac{T^0_{\gamma}}{m_\nu}(1+z)\\
    &\approx 151(1+z)\left(\frac{1\,\mathrm{eV}}{m_\nu}\right) \,\mathrm{km\,s^{-1}},
    \label{eqn:thermal}
\end{align}
where $m_\nu$ is the neutrino eigenstate mass in $\mathrm{eV}$ and $T^0_\gamma=2.725\,\mathrm{K}$ is the photon temperature today. Then, the free-streaming scale for the non-relativistic neutrinos becomes
\begin{equation}
    k_{\rm fs}(z) \approx 0.81 \frac{\sqrt{1-\om+\om(1+z)^3}}{(1+z)^2}\left(\frac{m_\nu}{1\,\mathrm{eV}}\right)\, h \,\mathrm{Mpc}^{-1},
    \label{eqn:free_stream_wave2}
\end{equation}
with $h=H_0/(100\,\mathrm{km\,s^{-1}\,Mpc^{-1}})$.

The redshift of the transition between relativistic and non-relativistic regimes is
\begin{equation}
    1+z_{\rm nr} \approx 1980 \left(\frac{m_\nu}{1\,\mathrm{eV}}\right),
    \label{eqn:trans}
\end{equation}
after which the free-streaming scale starts to grow, since $k_{\rm fs}\propto (1+z)^{-1/2}$, passing a minimum corresponding to
\begin{equation}
    k_{\rm nr} \approx 0.018 \left(\frac{m_\nu}{1\,\mathrm{eV}}\right)^{1/2}\sqrt{\om}\, h \,\mathrm{Mpc}^{-1}.
    \label{eqn:free_stream_nr}
\end{equation}

Modes with $k>k_{\rm fs}$ result in the suppression of the growth of the cold dark matter (CDM) perturbations due to the weakening of the gravitational potential wells, whilst for $k<k_{\rm fs}$ perturbations are free to grow again. Free-streaming never affects modes with $k<k_{\rm nr}$, and the neutrino fluctuations evolve similarly to the CDM ones since the two fields are coupled. Nonetheless, the baryon perturbations remain suppressed and are free to grow in amplitude only after the matter-radiation decoupling, falling on the already formed gravitational neutrino damped CDM potentials. Thus, the galaxy bias, $b$, which is the amplitude of the matter clustering ought to be properly modelled accounting for scale dependence for studies concerning massive neutrinos. This is also true for the growth rate of structures, $f$, which is sensitive to neutrinos. 
 
 \begin{figure}
 \centering \includegraphics[width=\columnwidth]{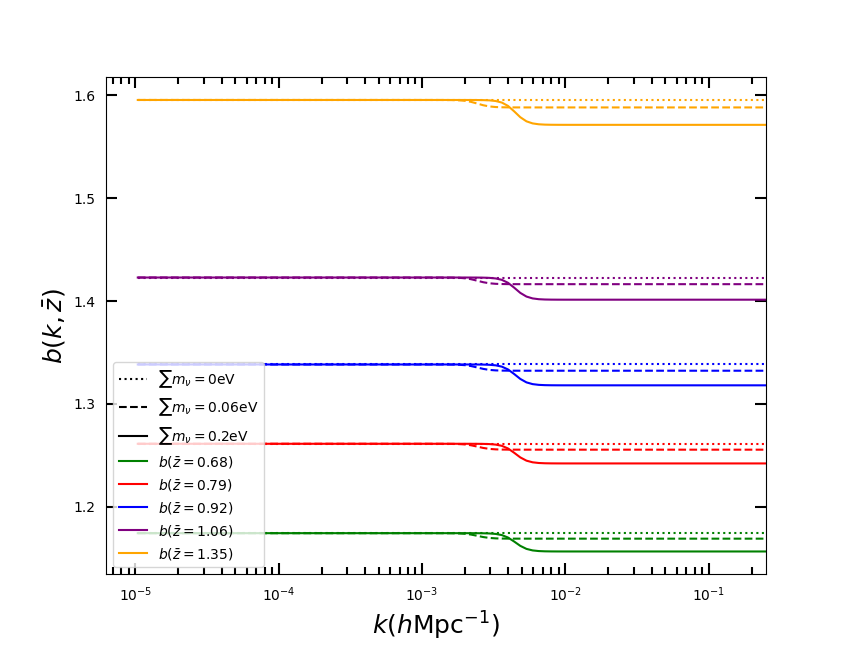}
     \caption{Scale dependent galaxy bias as a function of scale, with increasing redshift taken at the centre of each bin (see legend for colours). The scale-dependent bias is shown with dotted lines for $\sum m_\nu$=0eV, with dashed for $\sum m_\nu$=0.06eV and with solid for $\sum m_\nu$=0.2eV. The reference galaxy sample is the ELG catalogue.}
     \label{fig:scale_dep_bias}
 \end{figure}

For the scale dependent galaxy bias due to massive neutrinos, we use the recipe presented in \cite{Castorina_2014}. According to this,  the galaxy bias can be written with two definitions depending on the choice of the total matter or just the CDM and the baryon component, namely
\begin{align}
b^{\rm m}&=\sqrt{P^{\rm g}_{\rm lin}/P_{\rm lin}},\\
b^{\rm CDM+b}&=\sqrt{P^{\rm g}_{\rm lin}/P^{\rm CDM+b}_{\rm lin}},
\end{align}
with $P^{\rm g}_{\rm lin}$ and $P^{\rm CDM+b}_{\rm lin}$ the linear power spectra of clustering of galaxies and the CDM+baryon component, respectively. However, the galaxy formation is expected to be relevant for $k>k_{\rm fs}$, where neutrinos do not cluster. Thus, it is more precise to assume that the galaxies trace the field of the CDM+baryon perturbations and not the total matter field which includes neutrinos \citep{Vagnozzi_2018}. Hence, we opt for $b^{\rm CDM+b}$ as a definition of the galaxy bias.

In the case of $\Lambda$CDM cosmology with massive neutrinos and for $k\ll k_{\rm nr}$, the two definitions of $b^{\rm m}$ and $b^{\rm CDM+b}$ converge since the total matter power spectrum and the CDM+baryon power spectrum are the same, whilst for $k\gg k_{\rm nr}$ but well inside the linear regime we have the transition
\begin{equation}
    b^{\rm m} \rightarrow b^{\rm CDM+b}(1-f_\nu),
    \label{eqn:Castorina}
\end{equation}
where $f_\nu=\Omega_\nu/\om$ with $\Omega_\nu=\sum m_\nu/(93.14 h^2)$.

In order to account for a smooth transition for the linear galaxy bias values between $k\ll k_{\rm nr}$ and $k\gg k_{\rm nr}$, we use the expression
\begin{multline}
    b(k,z)=
    b_{k\ll k_{\rm nr}}(z)\\
    +\frac{b_{k\gg k_{\rm nr}}(z)-b_{k\ll k_{\rm nr}}(z)}{2}\left\{\tanh\left[\ln\left(\frac{k}{k_{\rm nr}}\right)^{\gamma}\right]+1\right\},
    \label{eqn:smooth_transition}
\end{multline}
with $b_{k\ll k_{\rm nr}}$ or $b_{k\gg k_{\rm nr}}$ the galaxy bias in the two asymptotic regimes and $\gamma$ setting the sharpness of the transition, for which we choose the value of 5---note that the actual value of $\gamma$ does not impact the results.
In \autoref{fig:scale_dep_bias}, we show the galaxy bias as a function of scale for different redshifts corresponding to the centres of each bin for the ELG galaxy sample (for colours, see legend). Solid, dashed and dotted lines are for $\sum m_\nu=0.2,\,0.06,\,0\,\mathrm{eV}$, respectively. It is evident that the amplitude of the transition and its position in scale is strongly dependent on the neutrino mass (see \autoref{eqn:free_stream_nr} and \autoref{eqn:Castorina}).

\section{Survey specifications}
\label{sec:thesurvey}
The coming decade will see a wealth of data from experimental campaigns aimed at scrutinising the large-scale structure of the Universe, such as the \textit{Euclid} satellite \citep{Laureijs2011,Amendola2013,Amendola2016}, the Rubin Observatory (formerly Large Synoptic Survey Telescope, LSST) \citep{2009arXiv0912.0201L}, the Square Kilometre Array (SKA) \citep{Maartens2015,Abdalla2015,SKA1_2018}, or the Dark Energy Spectroscopic Instrument (DESI) \citep{DESIcollab2016}.
Better to investigate the potentiality of multi-tracing, we choose to adopt as a reference experiment the last of those mentioned above, for reasons that will be clear in a moment.

\begin{figure}
\centering \includegraphics[width=\columnwidth]{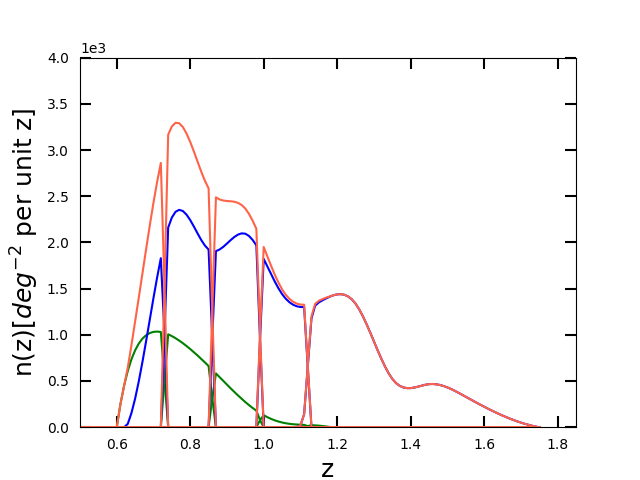}
    \caption{The ELG (blue) and LRG (green) galaxy sub-sample distributions in redshift, and their combination, for simplicity referred to as DESI (red).}
    \label{fig:DESI_bins}
\end{figure}

DESI is a ground-based large-scale structure experiment that has recently started its five year active period. It aims at measuring the baryon acoustic oscillations and the growth of structures via RSD. It will be a wide-area survey of $14,000\,\mathrm{deg}^2$ with a target list of galaxies and quasars observed spectroscopically. The target imaging objects are divided in three classes, thus providing us with different tracers within the same experiment. The very low redshift objects ($z<1$) will be the luminous red galaxies (LRG), whilst those in the intermediate redshifts ($1<z<1.7$) will be bright oxygen emission-line galaxies (ELG). Finally, at very high redshifts ($2.1<z<3.5$), quasars will be traced thanks to their neutral hydrogen distribution using the Ly-$\alpha$ forest absorption lines. Here, however, we will only consider the ELG and LRG galaxy sub-samples, and shall refer to the total DESI galaxy sample as their summed distribution. The ELG and LRG distributions as a function of redshift are presented in \autoref{fig:DESI_bins}. We divide both samples in such a way that edges of each bin coincide between the tracers, in order to fully exploit the overlap binning for the multi-tracer technique. Spedicifally, we consider four equi-spaced bins for each tracer, but note that the fifth and last bin, where the LRG number density almost vanishes whereas the ELG distribution extends to the end of the sample in redshift.

We take tabulated values of the number of galaxies $\de N^A$ residing in a given redshift range $\de z$ from \citet[][Table~2.3]{DESIcollab2016}. We then interpolate them to get the distribution of sources $n^A(z)=\de N^A/\de z$. At this point, since galaxies will be observed by DESI with a high redshift accuracy thanks to spectroscopic measurements, we choose a top-hat modelling for the binning, which we model as
\begin{equation}
   n_i^A(z)=\frac12\left[1-\tanh{\left(\frac{|z-\bar{z_i}|-\Delta z/2}{r \Delta z/2}\right)}\right],
\end{equation}
with $\bar{z_i}$ the $i$th bin centre, $\Delta z$ the bin width, and $r$ an edge smearing, which we chose to be $0.03$. The smearing of the bin edges ensures that the integration will be numerically stable.

Thus, inside bin $i$ of the $A$th galaxy population, $n_i^A(z)$ galaxies reside, and their total number counts is
\begin{equation}
    \bar n^A=\sum_{i=1}^{N^A_z} \bar n_i^A,
\end{equation}
whilst the angular number count of sources in each bin comes from
\begin{equation}
    \bar n_i^A=\int \de z\,n_i^A (z).
    \label{eqn:n_i_bin}
\end{equation}

The linear galaxy bias on scales $k\ll k_{\rm nr}$ for the ELG and LRG galaxy samples is given in \citet{DESIcollab2016} as
\begin{align}
    b_{\rm LRG}(z)&=1.70/D(k\ll k_{\rm nr},z),\\
    b_{\rm ELG}(z)&=0.84/D(k\ll k_{\rm nr},z).
\end{align}
Finally, for the combined DESI distribution we choose a weighted average, i.e.\
\begin{equation}
    b_{\rm DESI}(z)=\frac{n_{\rm LRG}(z)b_{\rm LRG}(z)+n_{\rm ELG}(z)b_{\rm ELG}(z)}{n_{\rm LRG}(z)+n_{\rm ELG}(z)}.
    \label{eq:weighted_avrg}
\end{equation}

\begin{table}
\centering
\caption{The lower and upper multipole cuts for the LRG, ELG galaxy samples, and their summed total DESI distribution. The $\ell_{\rm min}$ is set as the point where the relative error between the angular power spectra measurement of \cosmosis\ and \class\ is less than 5\%, whilst in the linear regime limit the upper cut is specified as $\ell_{\rm max}=\chi(\bar z_i) k_{\rm max}$, where $\bar z_i$ is the $i$th bin centre.}
\begin{tabular}{cccc}
\hline
\multicolumn{3}{c}{$\ell_{\rm min}$} & \multicolumn{1}{c}{$\ell_{\rm max}$} \\
\cline{1-3}
LRG & ELG & DESI & \\
\hline
\hline
$63$ & $124$ & $119$ & $611$ \\
$78$ & $82$ & $108$ & $705$ \\
$98$ & $99$ & $125$ & $791$ \\
$166$ & $133$ & $170$ & $872$ \\
$176$ & $20$ & $20$ & ($946$)$1075$ \\
\hline
\end{tabular}
\label{tab:mulcut}
\end{table}

\section{Mock data set and likelihood}
\label{sec:likelihood_scale_cuts}
Similarly to the previous pieces of work in this series of papers, we plan to forecast cosmological parameters with a Bayesian approach. For that purpose, we set up a Monte Carlo Markov Chain sampling of the parameter posterior in the cosmological+nuisance parameter hyperspace. We assume a Gaussian likelihood for the data and minimise the chi-square function
\begin{equation}
    \chi^2(\bm\theta) =\sum_{\ell,\ell^\prime=\ell_{\rm min}}^{\ell_{\rm max}} \left[{\bm d}^{AB}_\ell - {\bm t}^{AB}_\ell({\bm \theta})\right]^{\sf T}\left (\bm\Gamma^{AB}_{\ell\ell^\prime}\right)^{-1} \left[{\bm d}^{AB}_\ell - {\bm t}^{AB}_\ell({\bm \theta})\right],
    \label{eq:chi2}
\end{equation}
with `${\sf T}$' and `$-1$' denoting matrix transposition and inversion, respectively. In \autoref{eq:chi2}, ${\bm d}^{AB}_\ell=\{{\bm C}^{\rm g}_\ell(z_i^A,z_j^B)\}$ is the data vector, constructed for our fiducial cosmological model at $\bm {\theta}_{\rm fid}$ by flattening the $N_z^A\times N_z^B$ tomographic matrix in each of its $N_\ell$ multipole bins; ${\bm t}^{AB}_\ell({\bm \theta})$ is the corresponding theory vector. We remind the reader that $A,B$ label the galaxy sample, namely ELG or LRG. Note that other works have previously dealt with parameter estimation from multi-tracing with angular power spectra \citep[e.g.][]{Ferramacho:2014pua,Fonseca_2015,Fonseca:2016xvi,Gomes:2019ejy,bellomo2020beware}, but always with Fisher matrices. Here, we instead build up a proper data analysis pipeline, which we test with synthetic data.

For the data, we assume a Gaussian covariance matrix, which takes the signal input of \autoref{eq:Cl_TOT} and reads
\begin{multline}
\Gamma^{ij,A;mn,B}_{\ell\ell^\prime}=\frac{\delta^{\rm K}_{\ell\ell^\prime}}{2\ell\Delta\ell f_{\rm sky}}\\
\times\left[\widetilde C^{\rm g}_\ell(z^A_i,z^B_m)\widetilde C^{\rm g}_\ell(z^A_j,z^B_n)
+\widetilde C^{\rm g}_\ell(z^A_i,z^B_n)\widetilde C^{\rm g}_\ell(z^A_j,z^B_m)\right],\label{eq:covmat}
\end{multline}
where $f_{\rm sky}$ the sky fraction probed by the survey, $\Delta \ell$ the multipole bin width, and
\begin{equation}
\widetilde C^{\rm g}_\ell(z_i^A,z_j^B) = C^{\rm g}_\ell(z_i^A,z_j^B) + \frac{\delta^{\rm K}_{ij}\delta^{\rm K}_{AB}}{\bar n_i^A},
    \label{eq:noise}
\end{equation}
is the observed signal plus shot noise, with $\bar n_i^A$ defined in \autoref{eqn:n_i_bin}. In this analysis, we employ $N_\ell$=20 log-spaced multipole bins.

It is worth remarking that:
\begin{itemize}
    \item $A=B$ represents the two single-tracer cases of ELG-ELG and LRG-LRG auto-correlation power spectra;
    \item $A\ne B$ is the ELG-LRG cross-correlation;
    \item The multi-tracer case is obtained by flattening the data and theory vectors, as well as the covariance matrix, along the $A,B$ indexes, too, thus considering at the same time all auto- and cross-correlations between the different galaxy samples.
\end{itemize}

The multipole range where the angular power spectra of \autoref{eq:Cl_TOT} are calculated is comprised between a lower $\ell_{\rm min}$ and an upper $\ell_{\rm max}$ cut. The lower limit is set according to the Limber approximation, which holds for $\ell\gg1$. Following the same fashion of \citetalias{Tanidis:2019teo} and \citetalias{Tanidis:2019mag}, we compare our Limber approximated \cosmosis\ spectra with the full spectra provided by \class\ \citep{Lesgourgues2011,Blas2011,DiDio2013}, and hence set the $\ell_{\rm min}$ where the relative error between the two spectra is less than 5$\%$. This choice is justified since this difference is always inside the $1\sigma$ theory error bar of the signal measurement. The choice of the upper bound $\ell_{\rm max}$ is made due to the fact that we consider only the linear scales in our analysis. This cut is defined as $\ell_{\rm max}=\chi(\bar{z_i}) k_{\rm max}$, where $\bar z_i$ is the centre of each redshift bin and the maximum wavelength reads $k_{\rm max}=\pi/(2 R_{\rm min})$, with $R_{\rm min}$ the sphere radius inside which the over-density perturbations at present have a variance  given by
\begin{equation}
\sigma^2(R)=\frac{1}{2\pi^2}\int{\de k}\,k^2P_{\rm lin}(k)\left|\frac{3j_1(kR)}{kR}\right|^2,
\label{eq:highmul}
\end{equation}
and is set to $\sigma^2(R_{\rm min})=1$ corresponding the $k_{\rm max}=0.25\,h\,\mathrm{Mpc}^{-1}$. The lower and the upper multipole cuts (shown in \autoref{tab:mulcut}) are imposed in all the redshift bins of the LRG and ELG galaxy samples, as well as their summed total DESI distribution (see again \autoref{fig:DESI_bins}). Notice that the $\ell_{max}$ for the 5th bin of the LRG sample is different than that of the ELG and DESI, due to the lower $\bar z$ considered. This choice is reasonable since the LRG number density of galaxies in that bin goes quickly to zero as already explained in \autoref{sec:thesurvey}.
\begin{table*}
\centering
\caption{Cosmological and nuisance parameters fiducial values (\lcdm\ best-fit of \citealt{Ade2015}) with their priors. We consider on top of the \lcdm\ model the case of 1 massive and 2 massless neutrinos with a fiducial value set to the minimum mass $\sum m_\nu=0.06\,\mathrm{eV}$.}
\begin{tabularx}{\textwidth}{Xllll}
	\hline
	 Parameter description & Parameter symbol & Fiducial value & Prior type & Prior range \\
	\hline
	\hline
	Present-day fractional matter density & $\om$ & 0.3089 & Flat & $[0.1,0.6]$\\
	Dimensionless Hubble parameter & $h$ & 0.6774 & Flat & $[0.5,1.0]$\\
    Amplitude of clustering$^\ddag$ & $\sigma_8$ & 0.8159 & Flat & $[0.4,1.2]$\\
    Present-day (physical) fractional neutrino density & $\Omega_\nu h^2$ & 0.00064 ($\sum m_\nu=0.06\,\mathrm{eV}$) & Flat & $[0.00064,0.05]$\\
	\hline
    Present-day fractional baryon density & $\ob$ & 0.0486 & -- & --\\
    Slope of the primordial curvature power spectrum & $\ns$ & 0.9667 & -- & --\\
	Amplitude of the primordial curvature power spectrum$^\ddag$ & $\ln(10^{10}\As)$ & $3.064$ & -- & --\\
    Optical depth to reionisation & $\tau_{\rm re}$ & 0.066 & -- & --\\
	\hline
    Overall redshift range amplitude bias parameter$^\P$ & $\alpha$ & 1.0 & Flat & $[0.1,2.0]$ \\
	\hline
	 Per redshift bin amplitude bias parameter$^\S$ & $b^A_i$ & 1.0 & Flat  & $[0.1,2.0]$ \\
	 \hline
\end{tabularx}\label{tab:params}
\raggedright\footnotesize{$^\ddag$ Following the LSS convention, we choose to sample on the $\sigma_8$ parameter to account for the matter perturbations amplitude and not on the primordial amplitude $A_{\rm s}$.\\
$^\P$ Applied parameter prior range for the `realistic' scenario.\\
$^\S$ Applied parameter prior range for the `conservative' scenario .}
\end{table*}

We have checked that for the particular spectroscopic binning width choice the magnification bias correction is not affecting our analysis and therefore its contribution can be safely neglected (more about the importance of this effect on \citetalias{Tanidis:2019mag}). This is in agreement with the findings of \cite{jeliccizmek2020importance}, but note that in the case of wider bins, ignoring lensing magnification may lead to biased estimation of neutrino masses \citep{Cardona:2016qxn}.

\section{Results}
\label{sec:results}
To test the multi-tracer technique applied to harmonic-space power spectra and our data analysis pipeline, we compare the constraints on the cosmological parameter set $\bm \theta=\{\om,\,h,\,\sigma_8,\, \sum m_\nu \}$ (where we include 1 massive and 2 massless neutrinos) provided by the LRG and ELG galaxy sub-samples alone, the DESI total sample, and multi-tracing between LRG and ELG. We include nuisance parameters that need to be marginalised over, to account for the ignorance that we have on the galaxy bias. In this respect, we explore both of the following scenarios:
\begin{itemize}
    \item[$(i)$] A realistic case, with an overall normalisation nuisance parameter spanning the whole redshift range. 
    \item[$(ii)$] A conservative choice of a nuisance parameter per redshift bin.
\end{itemize}
It is worth noting that, as optimistic as the realistic choice may seem, \citetalias{Tanidis:2019teo} showed no substantial difference with the conservative choice in the final results, and therefore we deem this case worthy of investigation here.

For the forecasting analysis, we use the Bayesian sampler \emcee\ \citep{ForemanMackey2013}. For the four aforementioned galaxy samples, we construct the mock observables (data vector ${\bm d}^{AB}_\ell$ and covariance matrix given by \autoref{eq:covmat}) within a fiducial \lcdm+$\sum m_\nu$ with fixed scale dependent galaxy bias considering the RSD correction on the galaxy density field (the fiducial values are shown in \autoref{tab:params}). Then, we explore the parameter hyperspace of the set $\bm \theta$ along with the nuisance parameters until we reach convergence with the sampler (for the priors see again \autoref{tab:params}). In the analysis of \citetalias{Tanidis:2019teo} and \citetalias{Tanidis:2019mag} we studied the effect of ignoring galaxy clustering corrections like RSD and magnification bias on the estimated cosmological parameter set. Thus, the resulting posteriors were expected to be biased described by highly non-Gaussian and/or bimodal shapes due to the incomplete information in our modelling. For this reason, in the previous pieces of work we opted for the means. In the present analysis, however, we always fit the synthetic data with the same model (which is RSD on the density field and \lcdm+$\sum m_\nu$ cosmology) and therefore feel safe to opt for the peaks of the one-dimensional marginalised posterior.

\begin{figure*}
\centering 
\includegraphics[width=0.3\textwidth]{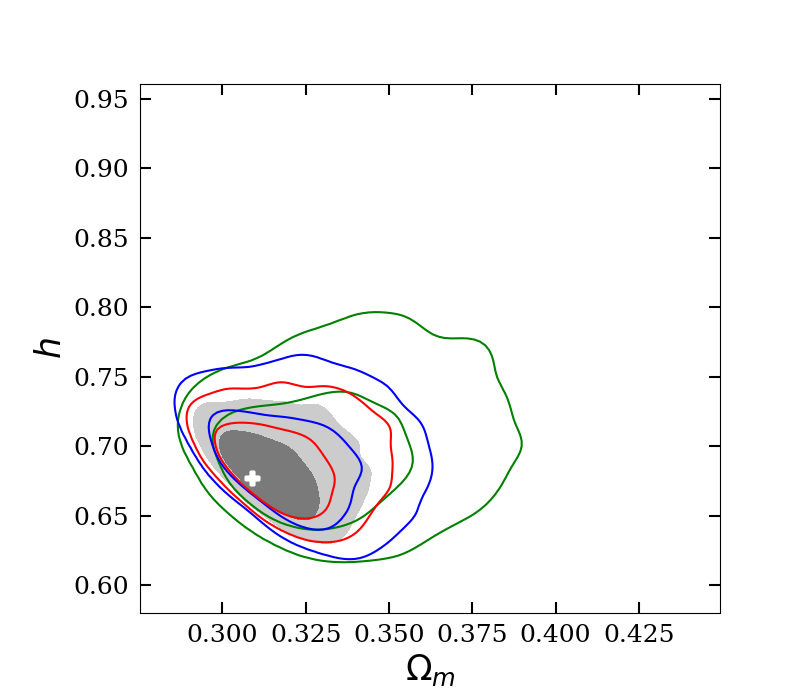}
\includegraphics[width=0.3\textwidth]{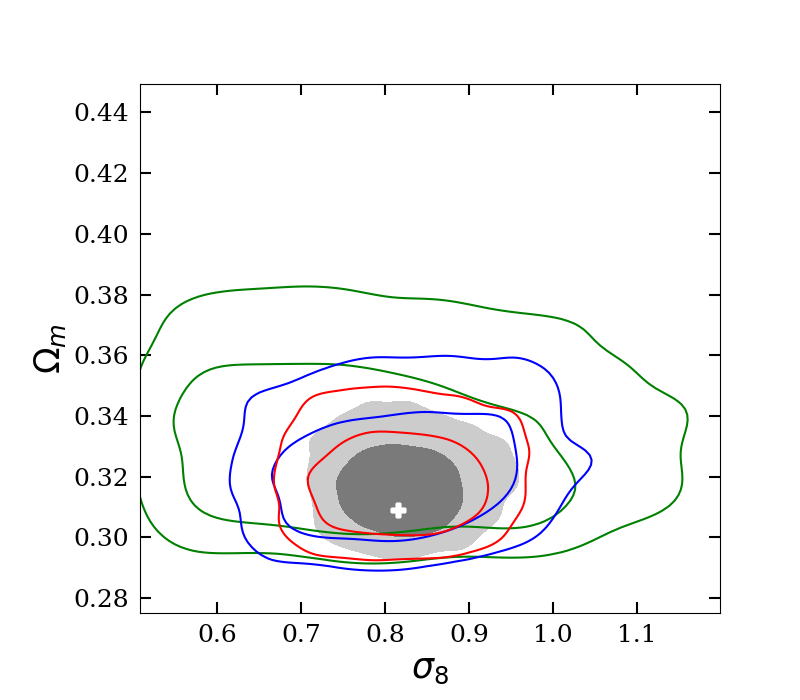}
\includegraphics[width=0.3\textwidth]{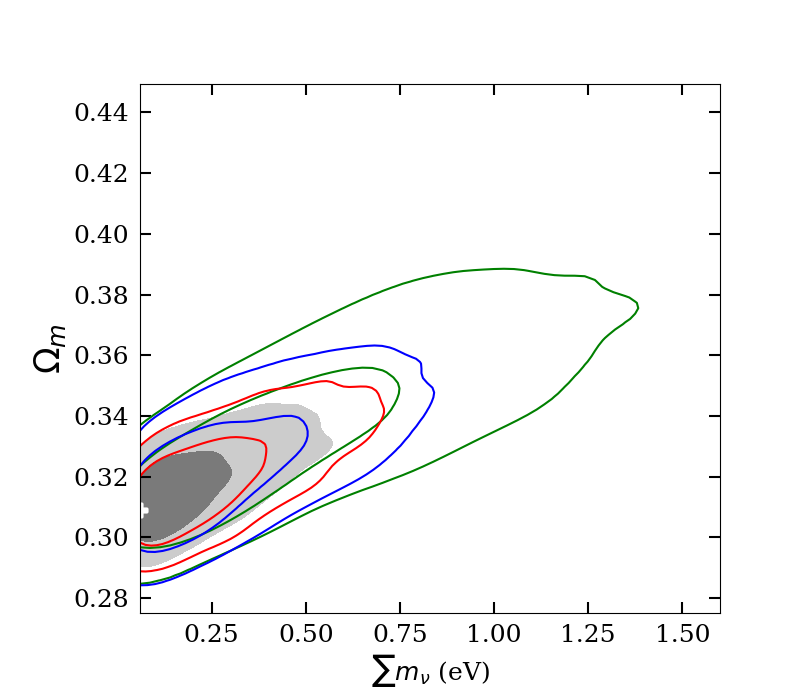}
    \label{fig:ConfidenceRegions_global_bias}
    \caption{Cosmological parameter constraints for the realistic galaxy bias case relation. The outer and inner contours are respectively the $95\%$ and $68\%$ C.L.\ on the marginal joint two-dimensional parameter space. The green, blue, and red empty contours correspond to the LRG, ELG, and DESI whilst the multi-tracer is presented with grey filled contours. The white cross stands for the fiducial cosmology considered for the generation of the synthetic data.}
\end{figure*}

\subsection{Realistic case}
In \autoref{fig:ConfidenceRegions_global_bias}, we present the $68\%$ and $95\%$ joint two-dimensional marginal confidence levels (C.L.) on the parameter set $\{\om,h,\sigma_8,\sum m_\nu \}$ for LRG (green), ELG (blue), DESI (red), and multi-tracing (grey), in the realistic case with the galaxy bias as given by \autoref{eqn:Castorina}. We keep the same colour code throughout the paper. The fiducial values used to generate the mock data and the priors for the Bayesian analysis are shown in \autoref{tab:params}, and as said we always include the RSD correction on the galaxy density field. It is worth mentioning that the contour on the $\om$-$h$ plane is quite skewed on the $\om$ direction. This is natural, since the neutrino fraction in the case of massive neutrinos is included in the total matter component and the uncertainty on the upper bound of their mass allows a higher $\om$ as well.  

In detail, we see that constraints on the parameters of interest, namely the $\om$, $h$, $\sigma_8$, and $\sum m_\nu$, from LRG are the weakest. The ELG distribution yields tighter constraints on the all the aforementioned parameters. This is expected, since the ELG has higher galaxy number density and the sample extents to a higher redshift range---note that the fifth bin for LRG is almost empty---containing in this way more cosmological information. The DESI combined distribution, as we have already mentioned, is the ELG and the LRG summed number density distribution with a weighted average galaxy bias given by \autoref{eq:weighted_avrg}, and naturally yields better results on the whole parameter set than the two separate samples.

Finally, when we consider multi-tracing between LRG and ELG, we get even tighter constraints, particularly on the $2\sigma$ upper bound on the sum of the neutrino masses. This effectively means an enhancement of $24\%$ with respect to the DESI bound. This is a major point in our analysis, attesting that with the multi-tracer technique we are able to considerably improve the results on the sum of the neutrino masses.

In addition, the more precise measurement on the scale dependent galaxy bias thanks to multi-tracing is clear by looking at the left panel of \autoref{fig:bias_plots}, where we show the peak of the marginalised one-dimension posterior and the $68\%$ error on the normalisation galaxy bias parameter. Here, we can appreciate that the multi-tracing (black) yields better constraints compared to those of the summed DESI galaxy distribution (red) by 30\%. 
% \stefc{This can be quantified---how much is that, $30\%$?} \textcolor{red}{[Wow, Stef you are amazing! you measured 30\% by eye?!]}\stefc{Thanks :)}
A similar trend can also be noticed on the $\sigma_8$ parameter (central panel of \autoref{fig:ConfidenceRegions_global_bias}), which is the normalisation of the power spectrum and is generally known to be degenerate with the galaxy bias.

\begin{table}
\caption{Marginalised one-dimensional posterior peak values, along with their $68\%$ C.L.\ intervals on the cosmological parameter set $\{\om,h,\sigma_8\}$ and the $95\%$ C.L.\ upper bound on the sum of neutrino masses. Results are obtained for the realistic bias scenario considering the LRG, the ELG galaxy sub-samples, the DESI total galaxy distribution, and the multi-tracer technique combining them all.}
\centering
\begin{tabularx}{\columnwidth}{llllX}
\hline
\multicolumn{5}{c}{Realistic scenario} \\
\hline
& LRG & ELG & DESI & Multi-tracer \\
\hline
\hline
\\
$\om$ & $0.321 ^{+0.025} _{-0.015}$ & $0.321 ^{+0.014} _{-0.017}$ & $0.3134 ^{+0.0155} _{-0.0086}$ & $0.3166 ^{+0.0092} _{-0.0109}$ \\
\\
$h$ & $0.683 ^{+0.039} _{-0.026}$ & $0.685 ^{+0.032} _{-0.026}$ & $0.684 ^{+0.026} _{-0.020}$ & $0.680 ^{+0.022} _{-0.019}$ \\
\\
$\sigma_8$  & $0.73 ^{+0.15} _{-0.16}$  & $0.813 ^{+0.098} _{-0.099}$  & $0.787 ^{+0.100} _{-0.041}$  & $0.808 ^{+0.056} _{-0.041}$ \\
\\
$\sum m_\nu$ & $<0.980\,\mathrm{eV}$  & $<0.578\,\mathrm{eV}$  & $<0.486\,\mathrm{eV}$  & $<0.369\,\mathrm{eV}$ \\
\\
\hline
\end{tabularx}
\label{tab:CastorinaRealistic}
\end{table}

\subsection{Conservative case}
\begin{figure*}
\centering \includegraphics[width=0.3\textwidth]{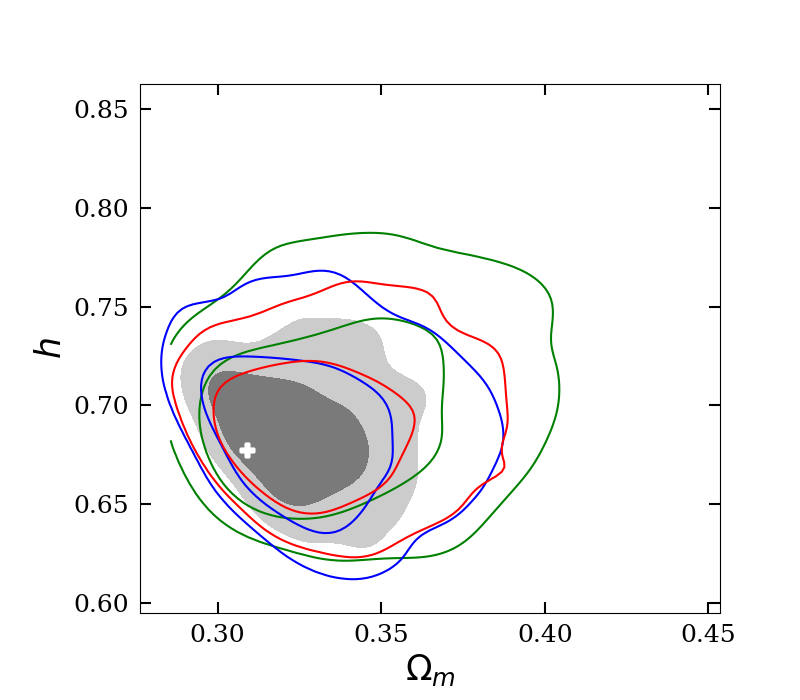}
\includegraphics[width=0.3\textwidth]{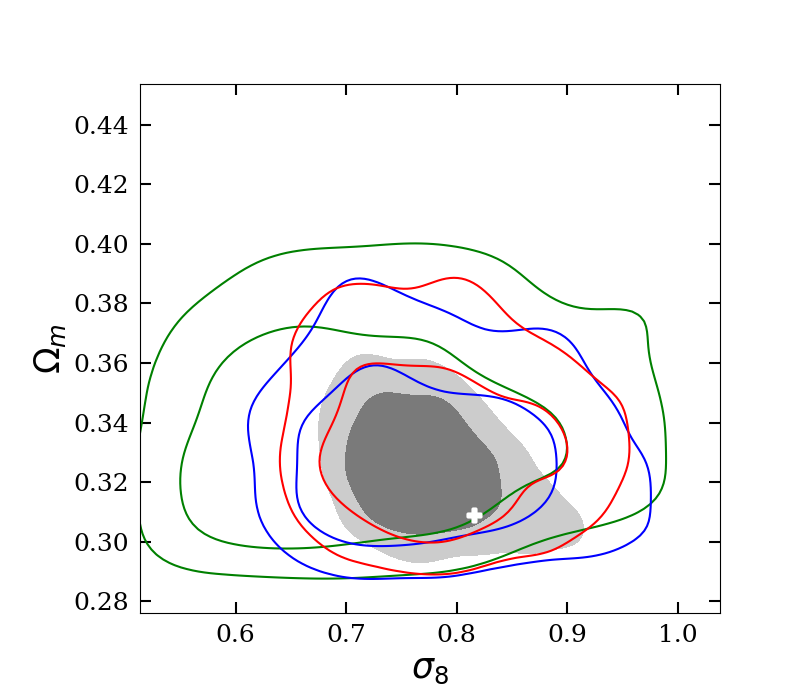}
\includegraphics[width=0.3\textwidth]{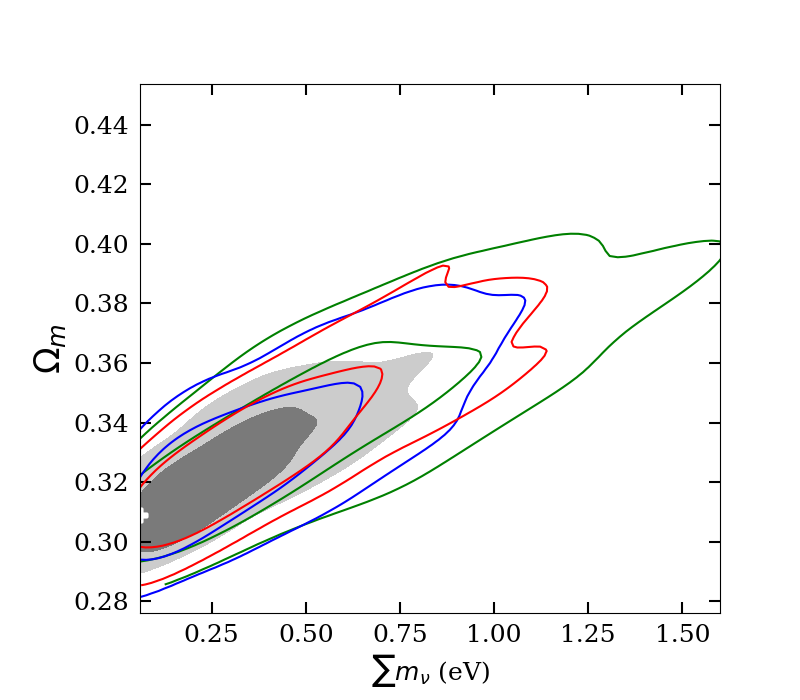}
    \caption{Same as \autoref{fig:ConfidenceRegions_global_bias} but for the conservative scenario}
    \label{fig:ConfidenceRegions_perbin_bias}
\end{figure*}

Similarly to the previous subsection, \autoref{fig:ConfidenceRegions_perbin_bias} presents the constraints on the cosmological parameter set of interest for all the galaxy samples, adopting a conservative scenario with a nuisance parameter per redshift bin ought to be marginalised over.

The results in the conservative case are quite similar to those obtained with the realistic scenario with the obvious exception that the upper bound on the sum of neutrino masses is weaker. This is a consequence of the fact that we have included more nuisance parameters in our modelling, and in particular galaxy bias parameters to which the neutrino masses are very sensitive, increasing in this way the measured error on this parameter. Generally, LRG yield again the weakest results, whilst better but comparable with each other are now the results obtained with ELG and DESI. This could also be attributed to the larger error bars due to the presence of nuisance parameters. The result of multi-tracing, however, are overall the most constraining again, having now a percentage gain of $44\%$ with respect to the DESI $2\sigma$ upper bound on the sum of the neutrino masses.

\begin{figure*}
\centering \includegraphics[width=0.45\textwidth]{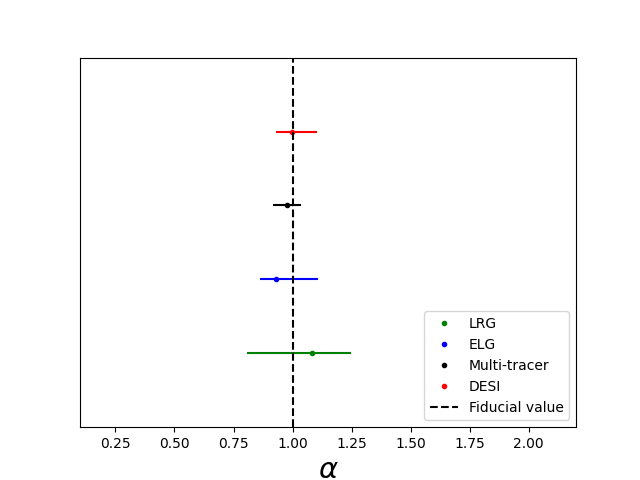}
\includegraphics[width=0.45\textwidth]{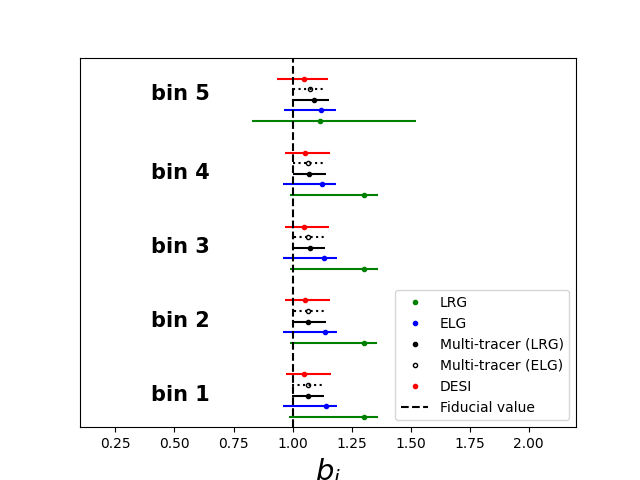}
    \caption{Marginalised one-dimensional peak values (bullets) and the $68\%$ C.L.\ asymmetric error bars (horizontal lines) for the realistic (left panel) and the conservative (right panel) amplitude galaxy bias parameters, considering the relation described by \autoref{eqn:Castorina}. We denote with green the LRG, with blue the ELG, with black the multi-tracer, and with red the DESI results. Note on the right panel the solid and dotted black error bars corresponding to the bias parameters of the first (LRG) and the second (ELG) tracer respectively in the multi-tracer technique. The vertical dashed black line stands for the fiducial cosmology value.}
    \label{fig:bias_plots}
\end{figure*}

Let us now focus on the constraints on the normalisation galaxy bias parameter per redshift bin, presented in the right panel of \autoref{fig:bias_plots}. Here, we can appreciate that the $68\%$ error bars obtained with the multi-tracing corresponding to either the first tracer LRG (green line) or the second tracer ELG (blue line) are tighter than the the DESI error bars (red line) by $\sim 30\%$. This holds also true for the normalisation of the power spectrum $\sigma_8$, as we can in the central panel of \autoref{fig:ConfidenceRegions_perbin_bias}. Finally, it is worth noting that the peak of the marginalised posterior of the normalisation galaxy bias values, although consistent within $68\%$ C.L.\ with the fiducial value, are slightly over-estimated. This counterbalances the peak of the one-dimensional marginalised posteriors of $\sigma_8$, which are oppositely a bit lower than the fiducial.

\begin{table}
\caption{Same as \autoref{tab:CastorinaRealistic} for the conservative bias scenario.}
\centering
\begin{tabularx}{\columnwidth}{llllX}
\hline
\multicolumn{5}{c}{Conservative scenario} \\
\hline
& LRG & ELG & DESI & Multi-tracer \\
\hline
\hline
\\
$\om$ & $0.316 ^{+0.040} _{-0.011}$ & $0.33 ^{+0.016} _{-0.023}$ & $0.323 ^{+0.023} _{-0.017}$ & $0.320 ^{+0.019} _{-0.014}$ \\
\\
$h$ & $0.687 ^{+0.037} _{-0.029}$ & $0.684 ^{+0.028} _{-0.023}$ & $0.684 ^{+0.026} _{-0.020}$ & $0.683 ^{+0.027} _{-0.018}$ \\
\\
$\sigma_8$  & $0.650 ^{+0.167} _{-0.054}$  & $0.717 ^{+0.120} _{-0.035}$  & $0.781 ^{+0.051} _{-0.085}$  & $0.766 ^{+0.051} _{-0.042}$ \\
\\
$\sum m_\nu$ & $<1.13\,\mathrm{eV}$  & $<0.716\,\mathrm{eV}$  & $<0.747\,\mathrm{eV}$  & $<0.420\,\mathrm{eV}$ \\
\\
\hline
\end{tabularx}
\label{tab:CastorinaConservative}
\end{table}

\section{Conclusions}
\label{sec:conclusions}
In this paper we have implemented the multi-tracer technique with a likelihood-based approach for harmonic-space power spectra of galaxy number counts, and we have investigated the potential of improving the $95\%$ C.L.\ upper bound on the sum of neutrino masses within a flat \lcdm$+\sum m_\nu$ model. For that purpose, we have considered the luminous red galaxy (LRG) and the emission-line galaxy (ELG) samples as envisaged to be measured by the Dark Energy Spectroscopic Instrument (DESI). We have conducted a synthetic data fitting with the \emcee\ sampler, and we included the redshift-space (RSD) correction to the spectrum of galaxy number density fluctuations. Since the contribution of massive neutrinos induces a scale dependence on the growth factor and the growth rate of structures, so does on the linear galaxy bias. For that purpose we have examined the scale dependence in the galaxy bias as described in \cite{Castorina_2014}.

The produced angular power spectra are in the Limber approximation and in the linear regime, so we have applied multipole cuts where the scales under consideration are valid. We did so for the case of four galaxy distributions. The LRG and ELG galaxy sub-samples, the total DESI distribution, and their multi-tracing, with a weighed average galaxy bias. Then, we have adopted two realistic scenarios to account for the ignorance on the galaxy bias with the introduction of nuisance parameters that should be marginalised over, namely: $i)$ an overall normalisation parameter, spanning the whole redshift range; and $ii)$ a conservative case with a nuisance parameter per bin.

We can summarise our results as follows:
\begin{itemize}
\item In the realistic scenario, the results obtained with the multi-tracing are overall the strongest with an enhancement of $24\%$ on the upper $95\%$ C.L.\ bound of the sum of neutrino masses with respect to the full DESI galaxy catalogue. Also, the $68\%$ C.L.\ errors on the galaxy bias nuisance parameter and on the normalisation of the matter power spectrum, $\sigma8$, are tighter by $30\%$ with multi-tracing.
\item The results of the conservative scenario are comparable with those of the realistic one. An expected difference is the weakening of the upper bound on the sum of the neutrino masses, due to the larger parameter space of the posterior. Still, multi-tracing yields significant improvement, with the $95\%$ C.L.\ upper bound of the sum of the neutrino masses $44\%$ more constraining than in the single-tracer case of the full DESI galaxy catalogue. We see, again, the same $\sim 30\%$ improvement on the $68\%$ C.L.\ error bar on the nuisance bias parameters and $\sigma_8$. 
\end{itemize}

Our results demonstrate the effectiveness of a harmonic-space analysis of multiple tracers, and represent a further improvement in our data analysis pipeline, described in \citetalias{Tanidis:2019teo} and \citetalias{Tanidis:2019mag}.

\section*{Acknowledgements}
We warmly thank Tamara Davis and Luis Raul Abramo for their invaluable comments that helped us to improve the presentation of our results. SC and KT acknowledge support from the `Departments of Excellence 2018-2022' Grant (L.\ 232/2016) awarded by the Italian Ministry of Education, University and Research (\textsc{miur}). SC also acknowledges support by \textsc{miur} Rita Levi Montalcini project `\textsc{prometheus} -- Probing and Relating Observables with Multi-wavelength Experiments To Help Enlightening the Universe's Structure' for the early stages of this project.

%%%%%%%%%%%%%%%%%%%%%%%%%%%%%%%%%%%%%%%%%%%%%%%%%%



\bibliographystyle{mnras}
\bibliography{Bibliography} % if your bibtex file is called example.bib

%%%%%%%%%%%%%%%%%%%%%%%%%%%%%%%%%%%%%%%%%%%%%%%%%%

%%%%%%%%%%%%%%%%% APPENDICES %%%%%%%%%%%%%%%%%%%%%

\appendix

\section{Our public code}
\label{appendix}

Our public code can be downloaded from the github repository \url{https://github.com/ktanidis/Modified_CosmoSIS_for_galaxy_number_count_angular_power_spectra}. This is a modified version of the publicly available code \href{https://bitbucket.org/joezuntz/cosmosis/wiki}{\cosmosis}. In this version, the galaxy number count angular power spectra are calculated in the linear regime, and the code allows for single and multiple tracers and for the inclusion of standard density fluctuations, RSD, and magnification bias. In the repository, we provide installation instructions in the \texttt{README.md} file. Specifically, we describe the required package dependencies and the set-up process. The physical framework of the code is described in \autoref{sec:formalism} of this paper. 

To account for single- or multi-tracer analysis, we use the \href{https://bitbucket.org/joezuntz/cosmosis/wiki}{\cosmosis} module \href{https://bitbucket.org/joezuntz/cosmosis/wiki/default_modules/load_nz_1}{load\_nz}. This module reads from a \texttt{.txt} file the distributions for the $i$th bin of the tracer $A$ with the format: $1$st column redshift $z$, and the remaining columns the $n_i^A(z)$ distribution in each bin. For example, for two tracers $A,B$ each having $2$ bins in the given redshift range, the columns read: \texttt{z},  \texttt{tracerA:bin1}, \texttt{tracerA:bin2}, \texttt{tracerB:bin1}, \texttt{tracerB:bin2}.

%\subsection{Module modifications}
The modified part of the code is the \texttt{project\_2d} module (for the original \href{https://bitbucket.org/joezuntz/cosmosis/wiki}{\cosmosis} module version see \href{https://bitbucket.org/joezuntz/cosmosis/wiki/default_modules/project_2d_1.0}{project\_2d}), and more specifically:
\begin{itemize}
    \item \texttt{utils.c}: Loads the function $P_{\,\mathrm{lin}}(k_\ell)$, and calculates $D(k_\ell,z)$, $f(k_\ell,z)$, $b(k_\ell,z)$ based on \autoref{eqn:Castorina};
    \item \texttt{kernel.c}: Specifies the considered galaxy number count contributions under the names \texttt{DEN} for the galaxy density field, \texttt{RSD} for redshift-space distortions and \texttt{MAG} for the weak lensing magnification. It also calculates the corresponding normalised $n_i^A(z)$ and some $\ell$-dependent prefactors;
    \item \texttt{limber.c}: Calculates the $C^{\rm g}_{\ell\gg1}(z^A_i,z^B_j)$ as in \autoref{eq:W_TOT}.
\end{itemize}

In addition to these, the python interface of the code is modified as well according to:
\begin{itemize}
    \item \texttt{limber.py}: Loads the source code functions;
    \item \texttt{project\_2d.py}: Provides the output. Three kernels are implemented with the names \texttt{W\_source}, \texttt{F\_source}, and \texttt{M\_source} accounting for \texttt{DEN}, \texttt{RSD}, and \texttt{MAG}, respectively.
\end{itemize}

\textsc{Note}: The current modified \cosmosis\ version is valid for galaxy clustering \textsc{only} under the entry \texttt{galcl-galcl=source-source-source} (see the example \texttt{.ini} file provided in the \texttt{README.md}). \textsc{Do not} attempt to ask output for the section names like \texttt{CMB\_kappa}, \texttt{Shear} or \texttt{Intrinsic alignments} etc.\ (for these see again the original module \href{https://bitbucket.org/joezuntz/cosmosis/wiki/default_modules/project_2d_1.0}{project\_2d}).

Finally, we modified the Gaussian likelihood module \href{https://bitbucket.org/joezuntz/cosmosis/wiki/default_modules/2pt_1}{2pt} to account for the output name \texttt{galcl}.

In order to test the code we provide an example \texttt{.ini} file (for more details see \texttt{README.md} and \texttt{.ini} file provided). The pipeline specified there gives the output mock data (data vector and covariance matrix) of multi-tracer galaxy number count angular spectra between the LRG and the ELG samples of the DESI \citep{DESIcollab2016} given a fiducial cosmological model. The cosmological parameter values are read from another \texttt{values.ini} file. In there, the user can also declare whether to include or not density fluctuations, RSD and magnification bias by specifying the value 1 or 0 respectively for the parameters \texttt{DEN}, \texttt{RSD} and \texttt{MAG} (default is 1).

%%%%%%%%%%%%%%%%%%%%%%%%%%%%%%%%%%%%%%%%%%%%%%%%%%

% Don't change these lines
\bsp	% typesetting comment
\label{lastpage}
\end{document}